\documentclass[sigconf,nonacm]{acmart}
\usepackage{breakurl}
\usepackage{color}
\usepackage{booktabs} 
\usepackage{listings}
\usepackage{balance}
\usepackage{amsmath}
\usepackage[mathscr]{euscript}
\usepackage{verbatim}

\begin{document}
\sloppy

\title{Parareal Algorithm Implementation and Simulation in Julia}

\author{Tyler M. Masthay}
\affiliation{%
  \institution{The Institute for Computational Engineering and Sciences}
  \institution{The University of Texas at Austin}
  \city{Austin}
  \state{Texas}
  \postcode{478712--1229~~USA}
}
\email{tyler@ices.utexas.edu}

\author{Saverio Perugini}
\orcid{0002-1736-4009}
\affiliation{%
  \institution{Department of Computer Science}
  \institution{University of Dayton}
  \streetaddress{300 College Park}
  \city{Dayton}
  \state{Ohio}
  \postcode{45469-2160}
}
\email{saverio@udayton.edu}

\renewcommand{\shortauthors}{T.M. Masthay \& S. Perugini}

\begin{abstract} We present a full implementation of the \textit{parareal
algorithm}---an integration technique to solve differential equations in
parallel---in the Julia programming language for a fully general, first-order,
initial-value problem.  We provide a brief overview of Julia---a concurrent
programming language for scientific computing.
Our implementation of the
\textit{parareal algorithm} accepts both coarse and fine integrators as
functional arguments.  We use \textit{Euler's method} and another
\textit{Runge-Kutta} integration technique as the integrators in our
experiments.  We also present a simulation of the algorithm for purposes of
pedagogy and as a tool for investigating the performance of the algorithm.
\end{abstract}


\maketitle

\section{Introduction}

The \textit{parareal algorithm} was first proposed in 2001 by Lions, Maday, and
Turinici~\cite{LionsEtAl} as an integration technique to solve differential
equations in parallel.  We present a full implementation of the parareal
algorithm in the Julia programming language
(\url{https://julialang.org})~\cite{Julia} for a fully general, first-order,
initial-value problem.  Furthermore, we present a simulation of the algorithm
for purposes of pedagogy and as a tool for investigating the performance of the
algorithm.  Our implementation accepts both coarse and fine integrators as
functional arguments.  We use \textit{Euler's method} and another
\textit{Runge-Kutta} integration technique as the integrators in our
experiments.  We start with a brief introduction to the Julia programming
language.

\section{An Introduction to Julia:
Dynamic, yet Efficient, Scientific/Numerical Programming}

Julia is a multi-paradigm language designed for scientific computing; it
supports multidimensional arrays, concurrency, and metaprogramming.  Due to
both Julia's LLVM-based Just-In-Time compiler and the language design, Julia
programs run computationally efficient---approaching and sometimes matching the
speed of languages like C.  See~\cite{website} for a graph depicting the
relative performance of Julia compared to other common languages for scientific
computing on a set of micro-benchmarks.

\subsection{Coroutines and Channels in Julia}

Coroutines are typically referred to as \textit{tasks} in Julia, and are not
scheduled to run on separate CPU cores.  Channels in Julia
can be either synchronous or asynchronous, and can be typed.  However,
if no type is specified in the definition of a channel, then values of any
type can be written to that channel, much like \textsc{unix} pipes.
Messages are passed between coroutines through channels with the \texttt{put!}
and \texttt{take!()} functions.
To add tasks to be automatically scheduled, use the \texttt{schedule()}
function, or the \texttt{@schedule} and \texttt{@sync} macros.
Of course, coroutines have little overhead, but will always run on the same
\textsc{cpu}.

\begin{quote}
The current version of Julia multiplexes all tasks onto a single \textsc{os} thread.
Thus, while tasks involving \textsc{i/o} operations benefit from parallel execution,
compute bound tasks are effectively executed sequentially on a single \textsc{os}
thread. Future versions of Julia may support scheduling of tasks on multiple
threads, in which case compute bound tasks will see benefits of parallel
execution too\cite{JuliaParallelComputingMan}.
\end{quote}

\subsection{Parallel Computing}

In addition to tasks, Julia supports
parallel computing---functions 
running on multiple \textsc{cpu}s or distributed computers.
New processes are spawned with
\texttt{addproc(}$\mathtt{<}$\texttt{\textit{n}}$\mathtt{>}$\texttt{)}, where
$\mathtt{<}$\texttt{\textit{n}}$\mathtt{>}$ is the number of processes desired.
The function \texttt{addproc} returns the \textsc{pid}s of the created
processes. The function \texttt{workers} returns a list of the processes.
Alternatively, the Julia interpreter can be started with the \texttt{-p}
$\mathtt{<}$\texttt{\textit{n}}$\mathtt{>}$ option, where
$\mathtt{<}$\texttt{\textit{n}}$\mathtt{>}$ is the number of processes desired.
For instance:

\begin{lstlisting}[language=Matlab]
$ julia
julia> addprocs(3)
3-element Array{Int64,1}:
 2
 3
 4

julia> workers()
3-element Array{Int64,1}:
 2
 3
 4

^D
$
$ julia -p 3
julia> workers()
3-element Array{Int64,1}:
 2
 3
 4

^D
$
\end{lstlisting}

\noindent Note that the process ids start at 2 because
the Julia REPL shell is process 1.

Processes in Julia, which are either locally running or
remotely distributed, communicate with each other through message passing.  

The function
\texttt{remotecall(}$\mathtt{<}$\texttt{\textit{Function}}$\mathtt{>}$\texttt{,}~$\mathtt{<}$\texttt{\textit{ProcessID}}$\mathtt{>}$\texttt{,}~$\mathtt{<}$\texttt{\textit{args}
\dots}$\mathtt{>}$\texttt{)} executes
$\mathtt{<}$\texttt{\textit{Function}}$\mathtt{>}$ on worker
$\mathtt{<}$\texttt{\textit{ProcessID}}$\mathtt{>}$ and returns a value of the
\texttt{Future} type, which contains a reference to a location from which the
return value can be retrieved, once
$\mathtt{<}$\texttt{\textit{Function}}$\mathtt{>}$ has completed its
execution.  The \texttt{Future} value can be
extracted with the function \texttt{fetch()}, which blocks until
the result is available.
Thus, the function \texttt{remotecall} is used to send a message while the
function \texttt{fetch} is used to receive a message.
For instance:

\begin{lstlisting}[language=Matlab]
julia> addprocs(2)
julia> future = remotecall(sqrt, 2, 4)
Future(2,1,3,Nullable{Any}())
julia> fetch(future)
2.0
\end{lstlisting}

\noindent
After the function \texttt{remotecall} is
run, the worker process simply waits for the next call to \texttt{remotecall}.

\begin{lstlisting}[language=Matlab]
julia> counter1 = new_counter(3)
(::#1) (generic function with 1 method)
julia> future = remotecall(counter1, 2)
Future(2,1,23,Nullable{Any}())
julia> fetch(future)
4
\end{lstlisting}

\noindent The Julia macro \texttt{@spawn} simplifies this message-passing
protocol for the programmer and
obviates the need for explicit use of the low-level \texttt{remotecall}
function.  Similarly, the macro \texttt{@parallel} can be used to run each
iteration of a (for) loop in its own process.

\begin{lstlisting}[language=Matlab]
julia> future = @spawn sqrt(4)
julia> fetch(future)
2.0
julia> addprocs(2)
2-element Array{Int64,1}:
 3
 4
julia> @everywhere function fib(n)
    if (n < 2)
        return n
    else 
        return fib(n-1) + fib(n-2)
    end
end
julia> @everywhere function fib_parallel(n)
    if (n < 35)
        return fib(n)
    else
        x = @spawn fib_parallel(n-1)
        y = fib_parallel(n-2)
        return fetch(x) + y
    end
end
julia> @time fib(42)
  2.271563 seconds (793 allocations: 40.718 KB)
267914296
julia> @time fib_parallel(42)
  3.483601 seconds (344.48 k allocations:
                    15.344 MB, 0.25% gc time)
\end{lstlisting}

\noindent
There are also remote channels which are writable for more control over
synchronizing processes.

\subsection{Multidimensional Arrays}

Julia supports multidimensional arrays, an important data structure
in scientific computing applications, with a simple syntax and their efficient
creation and interpretation over many dimensions~\cite{Operators:Julia}.  The
function call
\texttt{ArrayType(}$\mathtt{<}$\texttt{\textit{dimensions}}$\mathtt{>}$\texttt{)}
creates an array, where the $n^{\mathrm{th}}$ argument in
$\mathtt{<}$\texttt{\textit{dimensions}}$\mathtt{>}$ specifies
the size of the $n^{\mathrm{th}}$ dimension of the array.  Similarly, the
programmer manipulates these arrays using function calls that support
infinite-dimensional arrays given only limitations on computational time.

In summary, Julia incorporates concepts and mechanisms---particularly
concurrency and multidimensional arrays---which support efficient scientific
computing.

\section{The Parareal Algorithm}

\begin{figure}
\resizebox{\columnwidth}{!}{
\includegraphics[scale=1.0]{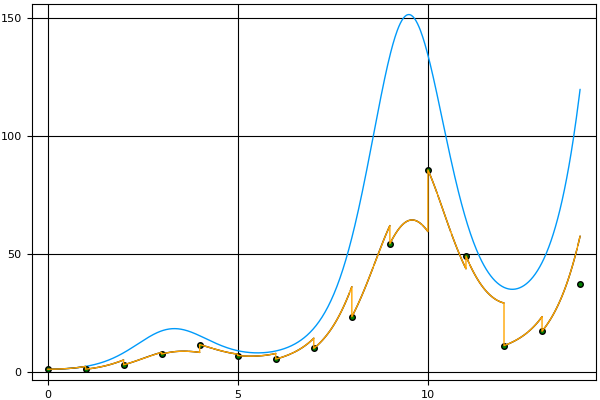}}
\caption{Right endpoint error.}
\label{errorIllustration}
\end{figure}

The parareal algorithm is designed to perform parallel-in-time integration for
a first-order initial-value problem.  The algorithm involves two integration
techniques, often known as the `coarse' integrator and the `fine'
integrator. For the algorithm to be effective, the coarse integrator must be of
substantially lower computational cost than the fine integrator. The reason
will become apparent later in this section.   Consider the differential
equation $\eqref{ode}$ given by

\begin{equation}\label{ode}
y'(t) = f(t,y(t)) \quad t \in [a,b]
\end{equation}

\noindent
with its associated initial-value problem $\eqref{ivp}$
\begin{equation}\label{ivp}
y(t^{*}) = y^{*} \quad t^{*} \in [a,b]\mathrm{.}
\end{equation}

\noindent For simplicity, let us assume $t^{*} = a$, so that the solution only
extends rightward.  To obtain an approximate solution to equation~$\eqref{ode}$
satisfying the initial condition~$\eqref{ivp}$, we partition our domain into
$[t_0=a,...,t_N=b]$ with uniform step size $\Delta$.
We now precisely define an `integrator' as a function from $(0,\infty) \times \mathbb{R}^{2} \times \mathscr{R}$ to $\mathbb{R}$ where $\mathscr{R}$ is the set of all Riemann integrable functions. For example, the integrator $I$ given by 
\begin{equation*}
I(\delta,x_0,y_0,g) = y_0 + g(x_0,y_0)\delta
\end{equation*}

\noindent
is the integrator corresponding to Euler's method with step size $\delta$.
Let $\mathscr{C}$ and $\mathscr{F}$ be the coarse and fine integrators, respectively. Define
\begin{align*}
   y_{0,1} &= y(t_0) = y^{*} \\
    y_{n+1,1} &= y(t_{n+1}) = \mathscr{C}(\Delta,t_n,y_{n,1},f)\mathrm{.}
\end{align*}

\noindent
Since $y_{n+1,1}$ depends on $y_{n,1}$, this algorithm is inherently
sequential. Partition $[t_n,t_{n+1}]$ into
$\{t_{n}^{0}=t_n,...,t_{n}^{m},...t_{n}^{M}=t_{n+1}\}$ with uniform step size $\delta < \Delta$. Define 
\begin{align*}\label{parallelDiscrete}
   z_{n,1}^{0} &= y(t_{n}^{0}) = y_{n,1} \\
    z_{n,1}^{m+1} &= y(t_{n}^{m+1}) = \mathscr{F}(\delta,t_{n}^{m},z_{n,1}^{m},f)\mathrm{.}
\end{align*}

\noindent
This yields an approximate solution $\{z_{n,1}^{0},...,z_{n,1}^{M}\}$ to
$\eqref{ode}$ over $[t_n,t_{n+1}]$ with initial conditions
\begin{align*}
   y(t_{n}) = y_{n,1}\mathrm{.}
\end{align*}

\noindent Since $z_{n_1,1}^{m_1}$ does not depend on $z_{n_2,1}^{m_2}$ for $n_1
\neq n_2$, we can compute these approximations in parallel. After the last
subproblem is solved, we simply combine the solutions on each subdomain to
obtain a solution over the whole interval. However, our values $\{
y_{1,1},...,y_{n,1} \}$ are relatively inaccurate. The vertical spikes in the
orange graph separating the coarse and fine predictions in
Figure~\ref{errorIllustration} illustrate this error.  However, $z_{n-1,1}^{M}$
is a better approximation for $\phi(t_n)$
 where $\phi$
is the exact solution to the differential equation.
We use this to obtain a 
better set of points $\{y_{n,2}\}$ for the coarse approximation. We do this by first defining $w_{n,1} = y_{n,1}$ and then defining
\begin{align*}
w_{1,2} &= y_{1,1} = y_{1,2} = y^{*} \\
w_{n,2} &= \mathscr{C}(\Delta,t_{n-1},y_{n-1,2},f)\\
y_{n,2} &= (w_{n,2}-w_{n,1}) + z_{n-1,1}^{M}\mathrm{.}
\end{align*}
Thus, $w_{n+1,2}$ serves as a new prediction given a more accurate previous prediction from $y_{n,2}$ since $z_{n-1,1}^{M}$ has now been taken into account in calculating $y_{n,2}$. In general, we continue evaluating so that for
$k > 1$, we have
\begin{align*}
w_{1,k} &= y_{1,k} = y^{*} \\
w_{n,k} &= \mathscr{C}(\Delta,t_{n-1},y_{n-1,k-1},f)\\
y_{n,k} &= (w_{n,k}-w_{n,k-1}) + z_{n-1,k-1}^{M}\mathrm{.}
\end{align*}

\noindent Note that since $y_{n,k}$ is dependent on $w_{n,k}$, this step must
be done sequentially. As $k$ increases, $w_{n,k} - w_{n,k-1} \to 0$, which
means that $y_{n,k}$ converges to the value that the fine integrator would
predict if fine integration were simply done sequentially. Thus, each $k$
denotes fine integration over the whole interval. This means that the total 
computation performed is much greater than if fine integration were performed
sequentially.  However, the time efficiency of each iteration has the potential
to be improved through concurrency. Since fine integration is more
computationally intensive, this improvement in the run-time efficiency may
compensate for the cumulative computation performed.

Let $K$ be the total number of iterations necessary to achieve a desired
accuracy of solution and $P$ be the number of subintervals into which we divide 
according to the coarse integrator. If $K=1$, then we achieve perfect parallel
efficiency. If $K = P$, then we likely slowed the computation down. The
parareal algorithm is guaranteed to converge to the solution given by the
sequential fine integrator within $P$ iterations. For a more complete treatment
of this convergence analysis, we refer the reader to~\cite{Gander}. For fully
general pseudocode, we refer the reader to~\cite{Aubanel,Nielsen}. 

\section{Parareal Algorithm Implementation in Julia}

\begin{lstlisting}[frame=tblr,basicstyle=\scriptsize,float=*,
caption={Implementation of the parareal algorithm in Julia.}, label=lst:euler]
@everywhere function parareal(a,b,nC,nF,K,y0,f,coarseIntegrator,fineIntegrator)
#initialize coarse information
xC = linspace(a,b,nC+1);
yC = zeros(size(xC,1),K);
deltaC = (b-a) / (nC + 1);
yC[1,:] = y0;

#"coarse integrator partially evaluated"
ciPEvaled = ((x1,y1) -> coarseIntegrator(deltaC,x1,y1,f));

#get initial coarse integration solution
for i=2:(nC+1)
   yC[i,1] = ciPEvaled(xC[i-1],yC[i-1,1]);
end
correctC = copy(yC);

#initialize fine information
xF = zeros(nC,nF+1);
for i=1:nC
   xF[i,:] = linspace(xC[i],xC[i+1],nF+1);
end
sub = zeros(nC,nF+1,K);
deltaF = xF[1,2] - xF[1,1];

#"fine integrator partially evaluated"
fiPEvaled = ((x1,y1) -> fineIntegrator(deltaF,x1,y1,f));

for k=2:K
   #run fine integration on each subdomain
   tic();
   @sync for i=1:nC
      sub[i,1,k] = correctC[i,k-1];
      @async for j=2:(nF+1)
         sub[i,j,k] = fiPEvaled(xF[i,j-1],sub[i,j-1,k]);
      end
   end
   toc();
   
   #predict and correct
   for i=1:nC
      yC[i+1,k] = ciPEvaled(xC[i],correctC[i,k]);
      correctC[i+1,k] = yC[i+1,k] - yC[i+1,k-1] + sub[i,nF+1,k];
   end
end

yF = zeros(nC*(nF+1),K-1);
for k=2:K
   yF[:,k-1] = reshape(sub[:,:,k]',nC*(nF+1));
end

return reshape(xF',nC*(nF+1)),reshape(sub[:,:,K]',nC*(nF+1)),yF,sub,xC,correctC,yC;
end
\end{lstlisting}

Listing~\ref{lst:euler} presents an implementation of the parareal algorithm
(from the prior section) in Julia. \noindent The \texttt{@async} macro within
the loop causes the program to evaluate the first expression to its right as a
concurrent task (i.e., the \texttt{for} loop assigning values to \texttt{sub}).
The \texttt{@sync} macro causes the main program thread to wait until all tasks
(spawned in the the first expression to its right with an \texttt{@async} or
\texttt{@parallel} macro) complete.  Once all concurrent tasks are complete,
execution of the program proceeds sequentially. Given the semantics of these
macros, the program in Listing~\ref{lst:euler}
correctly perform concurrent integration.
The sequential and parallel versions of this implementation have no significant
differences in run-time efficiency.  However, if a \texttt{sleep} statement is
placed in the argument of \texttt{fineIntegrator}, the parallel version runs
much faster.  This demonstrates that use of those two macros does lead to
concurrent program execution.

\section{Graphical Algorithm Simulation}

\begin{lstlisting}[frame=tblr,basicstyle=\scriptsize,float=*,
caption={Implementation of a graphical simulator of the parareal algorithm in Julia.}, label=lst:simulate]
@everywhere function fullMethod(n,a,b,y0,f,integrator)
   #setup domain and range space
    x = linspace(a,b,n+1);
   deltaX = x[2] - x[1];
    y = ones(n+1,1);
   
   #initialize left endpoint
    y[1] = y0;
   
   #integrate each point
    for i=1:n
        y[i+1] = integrator(deltaX,x[i],y[i],f);
    end
   return x,y;
end

function simulate(a,b,N,M,K,y0,f,coarseInt,fineInt,showPrev)
   x1,y1 = fullMethod(N*(M+1),a,b,y0,f,fineInt);
   x,y,yF,sub,xC,yC,iC = parareal(a,b,N,M,K,y0,f,coarseInt,fineInt);
   xF = (reshape(x,M+1,N))';
   fine = M+1;
   for k=2:K
      display(plot(x1,y1));
      if(showPrev && k > 2 )
         display(scatter!(xC,yC[:,k-2],color="red",legend=false));
      end
      display(scatter!(xC,yC[:,k-1],color="green",legend=false));
      done = zeros(Int64,N,1);
      workingSubdomains = 1:N;
      while(done != (M+1) * ones(N,1) )
         index = Int64(ceil(size(workingSubdomains,1)*rand()));
         currThread = workingSubdomains[index];
         while( done[currThread] == M+1 )
            currThread = Int64(ceil(N * rand()));
         end
         currThreadPlot = Int64(ceil(fine*rand()));
         totalAdvance = done[currThread] + currThreadPlot;
         if(totalAdvance > fine) totalAdvance = fine; end
         newP = (done[currThread]+1):totalAdvance;
         display(plot!(xF[currThread,newP],sub[currThread,newP,k],color="black"));
         done[currThread] = totalAdvance;
         workingSubdomains = find( ((x)->x != M+1), done );
         print(join(["Working on subdomain #", currThread, "...",
            "Pending Subdomains: ", workingSubdomains', "\n"]));
      end
      display(plot!(x,yF[:,k-1],color="orange"));
      sleep(5);
   end
end

# Implementation schemes.
function euler(delta,x0,y0,f)
   return y0 + delta * f(x0,y0);
end

function rungeKutta(delta,x0,y0,f)
   k1 = f(x0,y0);
   k2 = f(x0+delta/2,y0 + (delta/2)*k1);
   k3 = f(x0+delta/2,y0 + (delta/2)*k2);
   k4 = f(x0+delta,y0+delta*k3);
   return y0 + (delta/6)*(k1+2*k2+2*k3+k4);
end

\end{lstlisting}

The function \texttt{simulate} in Listing~\ref{lst:simulate} creates a
graphical simulator of the parareal algorithm.  This function can be used to
introduce the parareal algorithm to students in a numerical analysis course.
The first line gets the sequential solution from the fine integrator (the
`ideal' solution) and the second line gets the history of the computations that
took place during the parareal execution. The main loop over the variable $k$
then displays the inner workings of the algorithm. The ideal solution is
plotted, with a scatter plot of the points obtained from the coarse integrator.
To simulate the parallel nature of the algorithm, random progress is made on
randomly selected subdomains. Thus, the plot dynamically makes partial progress
on different subdomains until all subdomains are finished with the fine
integration. After this, the plots are connected into the current iteration's
approximation. During the next iteration, the previous guesses from the coarse
integrator are displayed in red and the new guesses from the coarse integrator
are displayed in green. As $k$ increases, these guesses converge to the ideal
solution.

\begin{figure*}
\begin{tabular}{cc}
\resizebox{\columnwidth}{!}{
\includegraphics[scale=1.0]{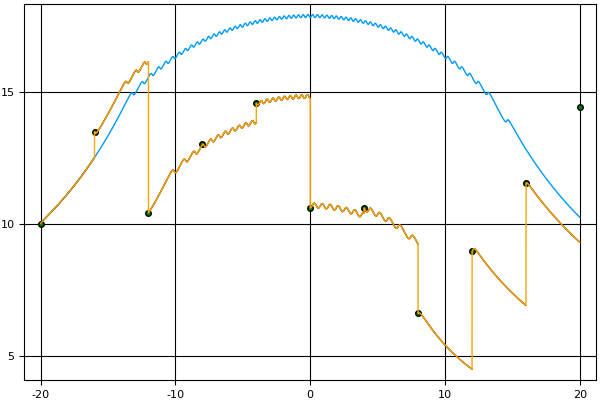}} &
\resizebox{\columnwidth}{!}{
\includegraphics[scale=1.0]{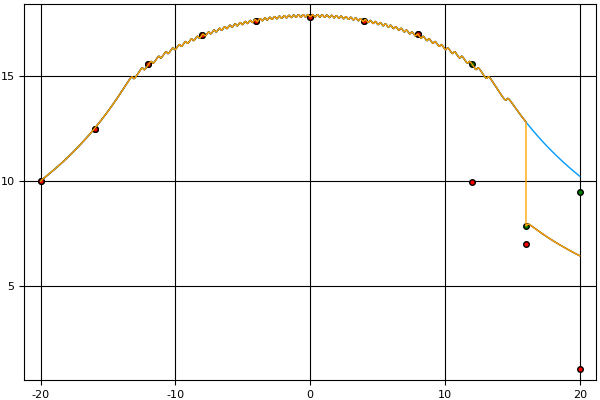}}
\end{tabular}
\caption{Slow parareal
example. (left) Solution after first iteration with Euler's method. (right)
Solution after ninth iteration with Euler's method.}
\label{sinPara}
\end{figure*}

In addition to the use of this function for pedagogical purposes, it can be
used to investigate the types of curves for which the parareal algorithm might
be practical.  For instance, consider the differential equation
\begin{equation*} y'(x) = sin(xy), \quad x \in [-20,20] \end{equation*}

\noindent with $y(-20) = 10$, $\Delta = 4$ (10 points), and $\delta = 0.008$
(500 points).  Figure~\ref{sinPara} shows the first and ninth iterations.  The
ninth iteration's large error on the right end of the interval shows that this
is an example where parareal convergence is slow. This is as inefficient as
possible, needing as many iterations as subdomains in order for the solution to
converge.  However, the simulation also shows that if $f(x,y) = sin(x)e^x$,
then the solution converges after just one iteration.  These two examples show
that the algorithm's efficiency can be highly dependent on the integrand. Below
the simulation function are \textit{Euler's method} and another
\textit{Runge-Kutta} integration technique that can be used as examples to be
passed as first-class functions as coarse or fine integration techniques to the
`parareal' or `simulate' functions.  A Git repository of both the
implementation and graphical simulation is available in BitBucket at
\url{https://bitbucket.org/sperugin/parareal-implementation-and-simulation-in-julia}.
All of the graphical plots are generated with the Julia Plots package available
at \url{https://juliaplots.github.io/}.  A video describing this application of
Julia is available on YouTube at
\url{https://www.youtube.com/watch?v=MtgbeLO6ZM4}.


\balance

\end{document}